\documentclass[a4paper,twocolumn,11pt,accepted=2024-07-14]{quantumarticle}
\pdfoutput=1
\usepackage{graphics,epsfig,color,mathrsfs,amsthm}
\usepackage{xcolor}% Include figure files
\usepackage{dcolumn}% Align table columns on decimal point
\usepackage{bm}% bold math
\usepackage[percent]{overpic}
\usepackage{physics}
\usepackage{mathtools}
\usepackage{hyperref}
\usepackage{booktabs}
\usepackage[normalem]{ulem}
\usepackage{wasysym}
\usepackage{verbatim}
\usepackage{caption}
\usepackage{amsmath,amssymb,physics}

\begin{document}
\title{Critical behaviors of non-stabilizerness in quantum spin chains}
%\title{}
\author{Poetri Sonya Tarabunga}
\affiliation{The Abdus Salam International Centre for Theoretical Physics (ICTP), Strada Costiera 11, 34151 Trieste,
Italy}
\affiliation{International School for Advanced Studies (SISSA), via Bonomea 265, 34136 Trieste, Italy}
\affiliation{INFN, Sezione di Trieste, Via Valerio 2, 34127 Trieste, Italy}

%\date{\today}

\begin{abstract}
Non-stabilizerness - commonly known as magic - measures the extent to which a quantum state deviates from stabilizer states and is a fundamental resource for achieving universal quantum computation.
In this work, we investigate the behavior of non-stabilizerness around criticality in quantum spin chains. To quantify non-stabilizerness, we employ a monotone called mana, based on the negativity of the discrete Wigner function. This measure captures non-stabilizerness for both pure and mixed states. We introduce Rényi generalizations of mana, which are also measures of non-stabilizerness for pure states, and utilize it to compute mana in large quantum systems.  We consider the three-state quantum Potts model and its non-integrable extension and we provide numerical evidence that the mutual mana exhibits universal logarithmic scaling with distance in conformal field theory, as is the case for entanglement. Comparing this with the scaling in gapped phases, we demonstrate that the scaling of mutual mana serves as a valuable tool for distinguishing between critical and non-critical behavior.

\end{abstract}
\maketitle

\section{Introduction}
 Over the last two decades, remarkable advancements in our understanding of many-body physics have been achieved through the exploration of concepts originating from quantum information theory and their application to quantum many-body systems. A prominent example is quantum entanglement \cite{horodecki2009quantum,vedral1997quantifying,nielsen2002quantum}, quantified by von Neumann and Rényi entropies, which has emerged as a powerful tool for investigating various many-body phenomena \cite{amico2008,eisert2010}, one of them is to identify universality classes in one-dimensional quantum critical points \cite{Holzhey1994,Calabrese2004}. 

Despite being a truly quantum property, it has been known that entanglement alone is insufficient to achieve universal quantum computation. Indeed, there exist states known as the stabilizer states that can be highly entangled, and yet they can be efficiently simulated on a classical computer \cite{gottesman1997stabilizer,gottesman1998theory,gottesman1998heisenberg,aaronson2004improved}. On the other hand, non-stabilizer states, often referred to as ``magic" states, play a fundamental role in realizing genuine quantum advantage \cite{bravyi2005UniversalQuantumComputation,bravyi2012magic,campbell2017roads,harrow2017quantum}. Non-stabilizer states are essential resources for achieving quantum computation beyond what classical systems can emulate. Much like entanglement, non-stabilizerness has been quantified within the framework of resource theory using measures of non-stabilizerness \cite{chitambar2019}. These measures assess the amount of resource a state can provide in quantum protocols involving only Clifford operations, offering insights into the computational power and quantum capabilities of different states.

In the many-body context, there have been several studies that suggest connection between non-stabilizerness and criticality \cite{white2021,Sarkar2020,oliviero2022ising, haug2023quantifying,tarabunga2023manybody,frau2024nonstabilizerness}.
At the same time, recent studies have also established that non-stabilizerness is directly linked with entanglement and Shannon (or participation) entropy \cite{tirrito2023,turkeshi2023measuring}. Specifically, it was found that the entanglement spectrum flatness (in any bipartition) \cite{tirrito2023} and participation entropy flatness \cite{turkeshi2023measuring} is directly related to the stabilizer linear entropy \cite{leone2022stabilizer}. Both mutual information (of entanglement) and Shannon mutual information have been shown to display the scaling relation \cite{Calabrese2004,Um2012,alcaraz2013,stephan2014, alcaraz2015,alcaraz2015_2,alcaraz2016}
\begin{equation} \label{eq:cft_scaling}
    I = \frac{c}{4} \log \left[ \frac{L}{\pi} \sin (\ell \frac{\pi}{L}) \right] + \gamma
\end{equation}
in critical spin chain governed by conformal field theory (CFT) on a periodic chain. Here $\ell$ is the subsytem size, $c$ is the central charge of the CFT, and $\gamma$ is a non-universal constant. Given the connections mentioned above, it becomes natural to question whether the corresponding mutual information of non-stabilizerness exhibits similar scaling behavior as in Eq. \eqref{eq:cft_scaling}. 

Addressing this question poses significant challenges. Firstly, directly evaluating non-stabilizerness becomes increasingly difficult for larger systems (especially since in principle Eq. \eqref{eq:cft_scaling} holds only for $\ell,L \gg 1$). Secondly, quantifying non-stabilizerness in mixed states, necessary for studying subsystems, is significantly more difficult compared to pure states. Previous studies on non-stabilizerness have been restricted to very small sizes \cite{white2021,Sarkar2020}, or relied on a non-stabilizerness monotone restricted to pure states \cite{haug2023quantifying,tarabunga2023manybody,haug2023stabilizer,tarabunga2023magic,frau2024nonstabilizerness,lópez2024exact}. To overcome these hurdles, this work focuses on quantum critical spin chains with odd on-site Hilbert space dimension. In such systems, there exists a strong measure of non-stabilizerness known as mana \cite{Veitch2012,Veitch2014}. Mana quantifies the non-stabilizerness for both pure states and mixed states, and its definition does not employ minimization procedures, making it the perfect choice to address the above question on scaling behavior of non-stabilizerness in critical systems. We leverage this advantage to investigate the behavior of mana in quantum critical spin chains governed by CFT. 

Prior investigations of mana have been limited to very small systems up to $L=6$ sites \cite{white2021,sewell2022}. This work significantly expands the capability of evaluating mana in substantially larger systems. To achieve this, we first introduce a R\'enyi generalization of mana, which we call mana entropies. These quantities are also measures of non-stabilizerness for pure states, although unlike mana they are not good measures for mixed states. We then construct a classical statistical mechanics systems derived from the discrete Wigner function, such that the computation of mana can be recast as a free energy calculation. We then show how this can be done by thermodynamics integration. Finally, we introduce the mutual mana and study its scaling in CFT. 

Our results demonstrate that the mana is significant at the critical point, and it exhibits a finite-size scaling. Moreover, we find that the mutual mana scales linearly with $\log \left[ \frac{L}{\pi} \sin (\ell \frac{\pi}{L}) \right]$, analogous to entanglement and Shannon entropy. Additionally, we show that the mutual mana instead saturates in gapped phases, thus showing the capability of the mutual mana to distinguish between critical and non-critical behavior.
 Our results highlight the difficulty of removing the non-stabilizerness in CFT with finite-depth quantum circuits, and in turn in classically simulating CFT.

The rest of the paper is structured as follows. In Sec.~\ref{sec:prelim}, we briefly cover some key preliminaries to provide the necessary background for introducing the non-stabilizerness monotone mana. In Sec.~\ref{sec:mana_entropy}, we introduce the Rényi generalizations of mana called the mana entropy, which themselves are new measures of non-stabilizerness for pure states, and in Sec.~\ref{sec:thermo} we present a thermodynamics view on the mana entropy, such that the computation of mana can be recast in the language of classical statistical mechanics. In Sec.~\ref{sec:mutual_mana}, we introduce the notion of mutual mana, and present a scalable method to compute them. In Sec.~\ref{sec:Pauli_Markov} we review the numerical method that we employ. In Sec.~\ref{sec:potts} we introduce the model under study and in Sec.~\ref{sec:num_results} we present our numerical results, both on the mana and mutual mana. Finally, we conclude in Sec.~\ref{sec:conclusions}.

\section{Preliminaries} \label{sec:prelim}
\subsection{Stabilizer formalism and resource theory of non-stabilizerness}
In this section, we review the stabilizer formalism for quantum systems of odd prime dimension and the resource theory of non-stabilizerness.
We first define the shift and clock operators as
\begin{equation} \label{eq:clock_and_shift}
X = \sum_{k=0}^{d-1} |k+1\rangle \langle k | \quad \textrm{and} \quad Z=\sum_{k=0}^{d-1} \omega^k_d |k\rangle \langle k |,
\end{equation}
 where $\omega_d=e^{2\pi i/d}$. Here, the addition is defined modulo $d$. They satisfy the commutation relation   $X Z = \omega_d Z X$.
 
 The generalized Pauli operators (also known as the Heisenberg-Weyl operators) are defined as
\begin{equation} \label{eq:pauli_qudit}
    T_{aa'} = \omega_d^{-2^{-1}aa'}Z^a X^{a'}
\end{equation}
for $a,a' \in \mathbb{Z}_d$. Here, $2^{-1}$ is the inverse element of $2$ in $\mathbb{Z}_d$. For a system of $N$ qudits, the Pauli strings are 
\begin{equation}
    T_{\mathbf{a}}=T_{a_1,a_1'} T_{a_2,a_2'} ...  T_{a_N,a_N'} .
\end{equation}
We denote the group of all $N$-qudit Pauli strings with phases as $\mathcal{P}_N$. 

The Clifford group $\mathcal{C}_N$ is defined as the normalizer of $\mathcal{P}_N$  
\begin{equation}
\mathcal{C}_N=\left \lbrace U   : \ UPU^{\dagger} \in \mathcal{P}_N,  \forall P \in \mathcal{P}_N \right \rbrace .
\end{equation}
The Clifford group can be generated using the qudit Hadamard gate, the phase gate, and the sum gate \cite{Hostens2005}. The pure stabilizer states are defined as all the states that can be generated by Clifford operations acting on the computational basis state $|0\rangle^{\otimes N}$. The set of stabilizer states is the convex hull of the set of pure stabilizer states:
\begin{equation}
\resizebox{.98\hsize}{!}{$
    \mathrm{STAB} = \left \lbrace \rho : \rho = \sum_j p_j \ket{S_j}\bra{S_j} , \forall j p_j \geq 0, \sum_j p_j = 1  \right \rbrace$},
\end{equation}
where $\ket{S_j}$ are pure stabilizer states for all $j$.
%On the other hand, universal quantum computation can be achieved through supplying magic (non-free) states. This can be achieved by augmenting the Clifford group with the Toffoli gate or the $\pi/8$-phase ($T$) gate, thus unlocking the potential for universal quantum computation.

It is well known by the Gottesman-Knill theorem  \cite{gottesman1997stabilizer,gottesman1998heisenberg,aaronson2004improved} that Clifford circuits and stabilizer states can be efficiently simulated on a classical computer. On the other hand, universal quantum computation can be achieved through supplying non-stabilizer states, thus
unlocking the potential for universal quantum computation \cite{bravyi2005UniversalQuantumComputation}. In this context, an important task is to quantify the amount of non-stabilizerness, which is measured using non-stabilizerness monotones in the framework of resource theories \cite{chitambar2019}. For systems with odd prime local dimension $d$, the resource theory has been developed \cite{Veitch2012,Veitch2014,Mari2012,Wang2020,Wang2019}. 

\subsection{Mana}
In this section, we introduce the non-stabilizerness monotone called mana \cite{Veitch2012,Veitch2014}, which we will employ in this work. Mana is a measure of non-stabilizerness that is only defined in terms of expectation values of operators, and is thus one of measures of non-stabilizerness that is relatively easy to compute. To define mana, we define the phase-space point operators in terms of the Pauli strings as 
\begin{equation} \label{eq:phase-space}
    A_{\mathbf{0}} = \frac{1}{d^N} \sum_\mathbf{u} T_\mathbf{u}, A_{\mathbf{u}} = T_{\mathbf{u}}  A_{\mathbf{0}} T_{\mathbf{u}}^\dagger.
\end{equation}
%These operators are Hermitian with eigenvalues $1$ and $-1$, with multiplicity $\frac{d+1}{2}$ and $\frac{d-1}{2}$, respectively. 
These operators are Hermitian with eigenvalues $1$ and $-1$. Moreover, they are orthogonal, i.e, $\Tr(A_\mathbf{a} A_\mathbf{b}) = d^N \delta(\mathbf{a},\mathbf{b})$, and thus they provide an orthogonal basis for an operator in $\mathbb{C}^{d^N \otimes d^N}$. Thus, one can expand the density matrix $\rho$ of a state (pure or mixed) as 
\begin{equation}
    \rho = \sum_\mathbf{u} W_{\rho} (\mathbf{u}) A_{\mathbf{u}}.
\end{equation}
where $W_{\rho} (\mathbf{u})$ is known as the discrete Wigner function \cite{Gross2006,Wootters1987}, a discrete analogue of the infinite-dimensional Wigner function \cite{wigner1932}. Equivalently, we can write
\begin{equation} \label{eq:wigner}
    W_{\rho}(\mathbf{u}) = \frac{1}{d^N} \Tr(A_{\mathbf{u}}\rho).
\end{equation}
The Wigner functions satisfy the following relations
\begin{subequations}
\begin{align}
 \sum_{\mathbf{u}} W_{\rho}(\mathbf{u})&=1\label{eq:normalization} \\
\sum_{\mathbf{u}} W_{\rho}(\mathbf{u})^2 &= e^{-S_2}/d^N\label{eq:purity}, 
\end{align}
\label{eq:wigner_relation}
\end{subequations}
where $S_2$ is the 2-Rényi entropy. 

Finally, mana is defined in terms of the Wigner functions as
\begin{equation} \label{eq:mana}
    \mathcal{M}(\rho) = \log (\sum_{\mathbf{u}}| W_{\rho}(\mathbf{u})|).
\end{equation}
Due to the normalization condition in Eq. \eqref{eq:normalization}, mana measures the negativity of the Wigner representation of $\rho$. For pure states, the set of states with positive Wigner representation is exactly the set of pure stabilizer states \cite{Gross2006}, in which case the mana vanishes. For mixed states, the set of states with positive Wigner representation is strictly larger than the convex hull of stabilizer states. Nevertheless, it is shown that states with positive Wigner representation (including those outside of the convex hull of stabilizer states) cannot be distilled \cite{Veitch2012}, and moreover they are efficiently simulatable \cite{pashayan2015}. In fact, mana directly quantifies the cost of classical simulation based on Monte Carlo in Ref. \cite{pashayan2015}. Thus, mana is a useful measure to quantify the resources required for classically simulating a quantum circuit, both for pure and mixed states \cite{Veitch2014}.

Crucially, mana stands out as the only known strong non-stabilizerness monotone whose definition bypasses the need for minimization procedures (Eq. \eqref{eq:mana}), both for pure and mixed states. This offers a significant computational advantage compared to other monotones. However, calculating mana still incurs an exponential cost as it necessitates computing the discrete Wigner function, $W_{\rho}(\mathbf{u})$, for all possible $\mathbf{u} \in \mathbb{Z}_d^N$. This exponential scaling renders direct calculation impractical for large systems, a key challenge that we address in this work.

\subsection{Stabilizer entropy} 
\label{sec:sre}
Stabilizer entropies (SEs) are a measure of nonstabilizerness recently introduced in Ref.~\cite{leone2022stabilizer}. For a pure quantum state $|\psi \rangle$ of a system of $N$ qudits, SEs are expressed in terms of the expectation values of all Pauli strings: 
\begin{equation} \label{eq:SRE_def}
M_n \left( |\psi \rangle \right)= \frac{1}{1-n} \log  \sum_{\mathbf{u}}  \frac{\left| \langle \psi | T_{\mathbf{u}} | \psi \rangle \right|^{2n}}{d^N}   \, .
\end{equation}
The definition for qubit systems is similar, by considering the Pauli operators for qubits.
Eq.~\eqref{eq:SRE_def} can be seen as the Rényi-$n$ entropy of the classical probability distribution $\Xi_{|\psi\rangle}(\mathbf{u})= \left| \langle \psi | T_{\mathbf{u}} | \psi \rangle \right|^2 / d^N$.
The SEs have the following properties~\cite{leone2022stabilizer,leone2024stabilizer}: (i) faithfulness, i.e., $M_n(|\psi \rangle)=0$ iff $|\psi \rangle \in \text{STAB}$; (ii) stability under Clifford unitaries $C \in \mathcal{C}_N$, i.e., $M_n(C|\psi \rangle )=M_n(|\psi \rangle)$; (iii) additivity, i.e., $M_n(|\psi \rangle_{A} \otimes|\psi \rangle_{B})=M_n(|\psi \rangle_{A})+M_n(|\psi \rangle_{B})$. Moreover, for qubits, it has been proven recently that the SEs are nonincreasing under stabilizer protocols $\mathcal{S}$ that map pure states to pure states, i.e, $M_n(\mathcal{E}(\ket{\psi}))  \leq M_n(\ket{\psi}), \forall \mathcal{E} \in \mathcal{S}$, for integer $n \geq 2$.

The key advantage of SE is its computability, with various numerical methods have been developed to compute the SEs ~\cite{haug2023stabilizer,lami2023quantum,tarabunga2023manybody,tarabunga2023magic,tarabunga2024nonstabilizerness}. However, unlike the mana, SEs lack the desirable property of strong monotonicity \cite{haug2023stabilizer}. Additionally, while a mixed-state extension of the SE has been proposed that retains the properties of the pure-state version \cite{leone2024stabilizer}, its computation requires minimization procedures, making it impractical to compute in large systems.

%Because the SE is defined in terms of the probability distribution $\Xi_{|\psi\rangle}(\mathbf{u})$, it is natural to estimate it through statistical sampling of $\Xi_{|\psi\rangle}(\mathbf{u})$. 

\section{Rényi generalizations of mana: mana entropy (ME)} \label{sec:mana_entropy}
In order to compute mana, we find it useful to introduce Rényi generalizations of mana, following closely the definition of SEs. We restrict to the case of pure states, where $\Pi_{|\psi\rangle}(\mathbf{u})= d^N W_{\rho}(\mathbf{u})^2$ can be interpreted as a probability distribution (see Eq. \eqref{eq:purity}), thus bearing similarity to $\Xi_{|\psi\rangle}(\mathbf{u})$. We now consider the $n$-Rényi entropies associated to this probability distribution in the same spirit as the SEs, as
\begin{equation}
\begin{split}
    \mathcal{M}_n &= \frac{1}{1-n} \log \sum_{\mathbf{u}} \left(d^N W_{\rho}(\mathbf{u})^2\right)^n - N \log d\\
     &= \frac{1}{1-n} \log \sum_{\mathbf{u}} \frac{\left| \Tilde{W}_{\rho}(\mathbf{u})\right|^{2n}}{d^N}   \\
    \end{split}
\end{equation}
where we define $ \Tilde{W}_{\rho}(\mathbf{u}):=d^N W_{\rho}(\mathbf{u})= \langle A_{\mathbf{u}} \rangle$.
Comparing this with Eq. \eqref{eq:SRE_def}, we see that  MEs are just SEs with the Pauli operators replaced by the phase-space point operators in Eq. \eqref{eq:phase-space}. It follows that the MEs possess similar properties as SEs, namely (i) faithfulness, (ii) stability under Clifford unitaries, and (iii) additivity. Moreover, they are upper bounded by $\mathcal{M}_n \leq N \log d$. 

Notice that the index $n=1/2$ corresponds to mana of pure states (up to a prefactor of 2). Mana has been rigorously proven to obey both monotonicity and strong monotonicity under stabilizer operations, making it a genuine measure of non-stabilizerness, also for mixed states \cite{Veitch2014}. In contrast, SEs of all index have been shown to violate strong monotonicity, while SEs of index $0<n<2$ violate monotonicity \cite{haug2023stabilizer}. It is presently unclear if such monotonicity property holds for MEs of index $n\neq 1/2$, a question that we leave for future investigations. Nonetheless, they could be useful to provide non-trivial bounds for other known measures of non-stabilizerness (see Appendix~\ref{sec:relations}). Moreover, while the computational cost to compute the mana grows exponentially in $N$, MEs of integer indices $n>1$ can be efficiently computed in matrix product states (MPS) with replica trick in the same way as SEs \cite{haug2023quantifying,tarabunga2024nonstabilizerness}. The same technique can also be used to obtain analytical results \cite{lópez2024exact}, which may be analytically continued to $n=1/2$ to obtain the mana. 

\subsection{Mana entropy and stabilizer entropy} \label{sec:mana_stabilizer}
Interestingly, we find that the mana entropy and stabilizer entropy is equivalent under some symmetry conditions, through the following proposition:

{\bf Proposition:} { \em Let $| \psi \rangle$ be an $N$-qudit pure state. If $A_\mathbf{b}$ is a phase-space operator such that $A_\mathbf{b} | \psi \rangle = \lambda | \psi \rangle$, where $\lambda \in \{ +1,-1\}$, then
\begin{equation}
    \lambda \langle \psi |  A_{\mathbf{a}+\mathbf{b}}   | \psi \rangle = \langle \psi |   T_{2\mathbf{a}}   | \psi \rangle \omega^{2(\mathbf{b}\mathbf{a'}-\mathbf{b'}\mathbf{a})}
\end{equation}
for all $\mathbf{a} \in \mathbb{Z}_d^{2N}$.
}

The proof can be found in Appendix~\ref{sec:proof}. As a corollary, the MEs and SEs are identical for all order whenever the state is stabilized by a phase-space operator (up to a sign). Importantly, we emphasize that mana entropies remain valid measures of non-stabilizerness even in cases where the equivalence with SEs does not hold.

\section{Thermodynamics approach to non-stabilizerness}
\label{sec:thermo}
We define a classical statistical system with energies $E_\mathbf{u}=-\log|\Tilde{W}_{\rho}(\mathbf{u})|$, such that the free energy is given by $F_\rho(\beta)=-\frac{1}{\beta}\log \sum_{\mathbf{u}} \left|\Tilde{W}_{\rho}(\mathbf{u})\right|^\beta$ \footnote{We note that similar thermodynamic description has been proposed for entanglement \cite{deBoer2019,yao2010,schliemann2011} and Shannon entropy \cite{zaletel2011}.}. One can see that the free energy is the same as the quantity $\frac{n-1}{2n} \mathcal{M}_n-\frac{N}{2n} \log d$ (for $n \neq 1$) with $\beta=2n$ \footnote{The case $n=1$ is instead related to the energy at $\beta=2$: $\mathcal{M}_1=2\langle E_\mathbf{u} \rangle_{\beta=2}$. As such, $\mathcal{M}_1$ can be directly estimated through perfect sampling techniques \cite{haug2023stabilizer,lami2023quantum} }. The calculation of $\mathcal{M}_n$ thus amounts to the computation of free energy of a classical system. Conventionally, this is commonly done by direct thermodynamics integration from infinite temperature ($\beta=0$). This is applicable when the free energy at infinite temperature is known, which is not generally true in this case. Luckily, the free energy at $\beta=2$ is known due to the relation in Eq. \eqref{eq:purity}. Indeed, for a pure state ($S_2=0$), Eq. \eqref{eq:purity} implies $F_\rho(\beta=2)=-\frac{N}{2} \log d$.  Thus, one can perform a direct thermodynamics integration starting from $\beta=2$,
\begin{equation} \label{eq:mana_thermo}
    \log \sum_{\mathbf{u}}\frac{|\Tilde{W}_{\rho}(\mathbf{u})|^\beta}{d^N} = \int_2^\beta \left\langle \log|\Tilde{W}_{\rho}(\mathbf{u})| \right\rangle_\beta d\beta,
\end{equation}
where $\langle ... \rangle_\beta$ denotes the thermal average at inverse temperature $\beta$. 

Numerically, the thermal average can be calculated via Monte Carlo sampling of the discrete Wigner function \cite{pashayan2015}. Here we perform the Monte Carlo sampling using tensor network methods, slightly modifying the method originally developed to compute SEs in Ref. \cite{tarabunga2023manybody}. In particular, we focus on mana, corresponding to $\beta=1$. 

\section{Mutual mana} \label{sec:mutual_mana}
We will also consider the ``mutual mana'' defined as
\begin{equation} \label{eq:mutual_mana}
    I_\mathcal{M}(A,B) = \mathcal{M}(\rho_{AB}) - \mathcal{M}(\rho_{A }) -\mathcal{M}(\rho_{B}).
\end{equation}
We will use the notation $ I_\mathcal{M}(\ell,L) $ to denote the case $A=  \{ 1,...,\ell \}$ and $B= \{ \ell+1,...,L \}$. Notice that the definition of mutual mana involves the mana of subsystems, which are mixed states.  Crucially, mana is a genuine measure of non-stabilizerness both for pure and mixed states, so that the mutual mana is a meaningful quantity that quantifies the amount of resource that resides in the correlations between parts of the system. It has also been suggested that it quantifies the difficulty of removing non-stabilizerness with a finite-depth circuit \cite{white2021}.

 We note here that mana is typically an extensive quantity. The subtraction in Eq. \eqref{eq:mutual_mana} thus serves to eliminate the leading extensive term, resulting in $I_\mathcal{M}(A,B)$ being significantly smaller than the mana itself. Extracting such a quantity through Monte Carlo samplings is known to be a challenging task, akin to the challenge of extracting topological entanglement entropy from entanglement entropy \cite{Isakov2011,block2020,Zhao2022}. Indeed, if one tries to compute $I_\mathcal{M}(A,B)$ by directly computing each of the three terms on the right hand side of Eq. \eqref{eq:mutual_mana} separately (e.g., using Eq. \eqref{eq:mana_thermo}), the resulting error bar will be prohibitively large. We overcome this difficulty by writing $I_\mathcal{M}(A,B)$ as  
\begin{equation} 
    I_\mathcal{M}(A,B) =  \log  \left( \frac{ \sum_{\mathbf{u},\mathbf{v}} |W_{\rho_{AB}}(\mathbf{u} \oplus \mathbf{v})| }{ \sum_{\mathbf{u}}|W_{\rho_{A}}(\mathbf{u})| \sum_{\mathbf{v}}|W_{\rho_{B}}(\mathbf{v})|}\right).
\end{equation}
In view of the thermodynamics description in the previous section, the expression inside the logarithm can be interpreted as a ratio of partition functions of the classical systems corresponding to $\rho_{AB}$ and $\rho_A \otimes \rho_B$. One way to estimate it in Monte Carlo simulations is by sampling from one classical system and averaging the ratio of the Boltzmann weights \footnote{We note that thermodynamics integration can also be employed in this case. However, this requires to compute the discrete Wigner function of both $\rho_{AB}$ and $\rho_A \otimes \rho_B$ at each Monte Carlo step. As the discrete Wigner function evaluation is computationally heavy, we choose the simpler technique described in the text, where we only need to sample the discrete Wigner function of $\rho_A \otimes \rho_B$.}.
 Concretely,  we consider the probability distribution  $\Pi_{\rho_{A(B)}}(\mathbf{u}) \propto |W_{\rho_{A(B)}}(\mathbf{u}))|$. We can estimate $I_\mathcal{M}(A,B)$ using 
\begin{equation} \label{eq:estimator_mutual_mana}
    I_\mathcal{M}(A,B) =  \log  \left\langle \frac{ |W_{\rho_{AB}}(\mathbf{u} \oplus \mathbf{v})| }{|W_{\rho_{A}}(\mathbf{u})||W_{\rho_{B}}(\mathbf{v})|}\right\rangle_{\Pi_{\rho_{A}}(\mathbf{u})\Pi_{\rho_{B}}(\mathbf{v})}.
\end{equation}

\begin{figure} 
    \centering
    \includegraphics[width=0.9\linewidth]{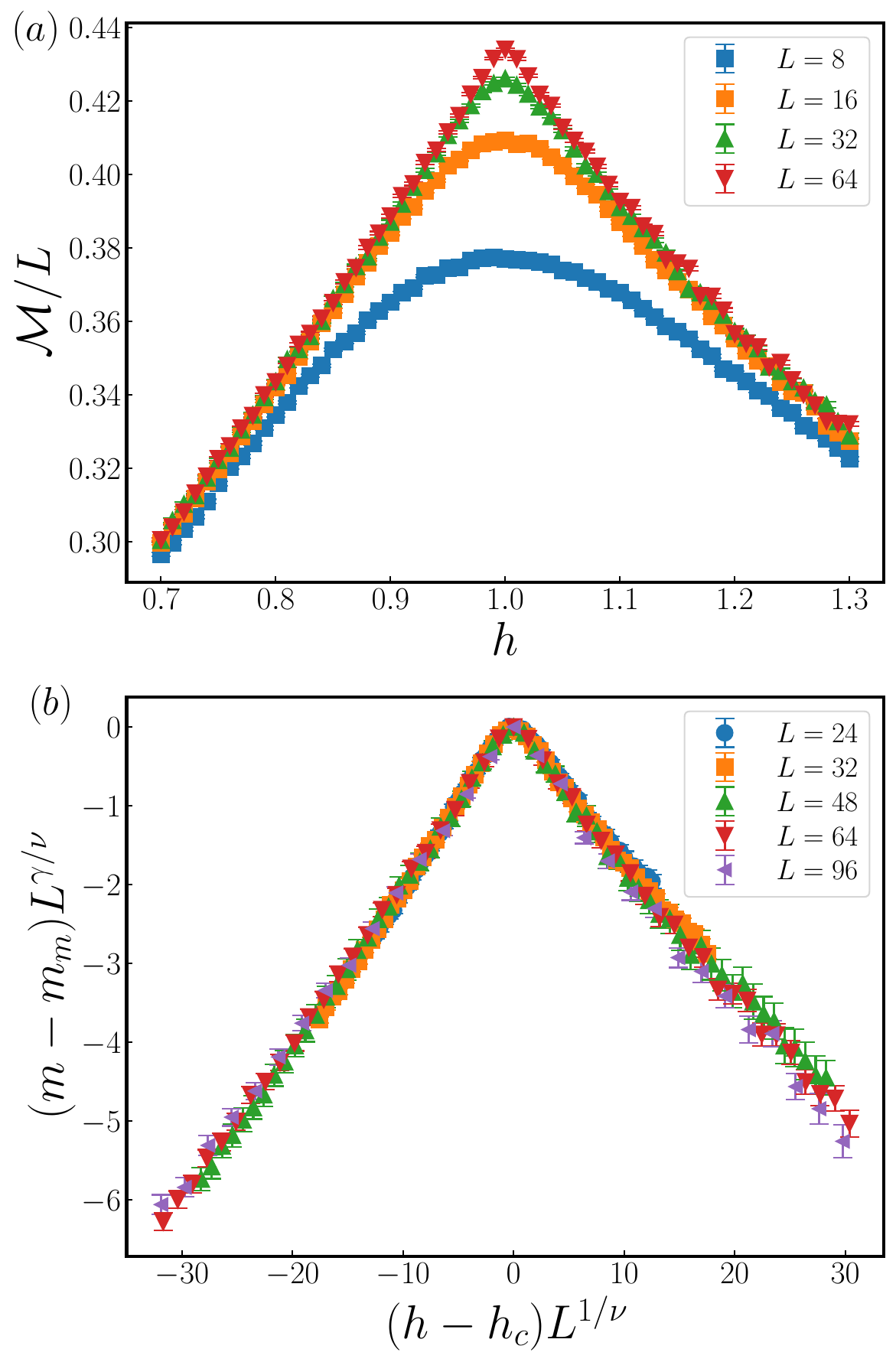}
    \caption{(a) Mana density $\mathcal{M}/L$ in the vicinity of the critical point $h=1$ in the three-state quantum Potts model. (b) Data collapse of the mana density $m=\mathcal{M}/L$ with $\gamma \approx 0.83$ and $\nu \approx 0.85$. The correlation-length exponent $\nu$ is close to the known $\nu_{Potts}=5/6$.}
    \label{fig:mana}
\end{figure}

\section{Numerical methods: Tree Tensor Network (TTN) sampling} \label{sec:Pauli_Markov}
In this section, we review the method introduced to estimate the SEs in Ref. \cite{tarabunga2023manybody}, which is based on sampling the Pauli strings using Monte Carlo scheme in tree tensor network (TTN). Here, we adapt it to instead sample the phase-space operators. With this technique, any probability distribution which only depends explicitly on expectation values can be sampled, thus enabling the calculation of both Eq. \eqref{eq:mana_thermo} and Eq. \eqref{eq:estimator_mutual_mana}. 

The core step is to compute the expectation value of any given phase-space operator, from which one can then perform the standard Metropolis algorithm to sample from the probability distribution of interest. Since phase-space operators are written as tensor products of single-site operators, their expectation values are efficiently computable with TTN (or any loopless tensor network \cite{Silvi2019}). Finally, from the samples of phase-space operators, one can then average over the estimators (for example, Eq. \eqref{eq:mana_thermo} and Eq. \eqref{eq:estimator_mutual_mana}) to estimate the quantities of interest.

To achieve efficient sampling within the TTN framework, we leverage its key property: any two tensors in the network are separated by at most $O(\log N)$ links. In this approach, the candidate phase-space operator for the next configuration only differs by a few sites (typically one or two) from the previous one. This enables the computation of the expectation value of the proposed operator in a highly efficient manner, requiring only $O(\log N)$ operations \cite{tarabunga2023manybody}. Crucially, the sites to be modified can be chosen arbitrarily, as long as the total number of changes does not scale with system size, which allows for flexible sampling strategies. The overall cost for each update scales as $O(\log (N) \chi^4)$, where $\chi$ is the bond dimension of the TTN.

The scheme can also be used to compute the mana of any partition of the system. To do this, we only need to restrict the phase-space operators to act only on the sites in the partition. The sampling procedure then proceeds as described above.

To emphasize the computational advantage of our newly proposed thermodynamics integration approach (Sec. \ref{sec:thermo}), it is worth recalling the estimators considered in Ref. \cite{tarabunga2023manybody}, that could be adapted to compute the mana entropies. If we sample according to $\Pi_{|\psi\rangle}(\mathbf{u})$, $\mathcal{M}_n$ can be estimated using the unbiased estimators
\begin{equation} \label{eq:estimator_n}
\mathcal{M}_n = \frac{1}{1-n} \log  \left\langle |\Tr (\rho A_{\mathbf{u}})|^{2(n-1)} \right\rangle_{\Pi_{|\psi\rangle}(\mathbf{u})} 
\end{equation}
for $n > 1$ and 
\begin{equation} \label{eq:estimator_1}
\mathcal{M}_1 =   \left\langle -\log \left( |\Tr(\rho A_{\mathbf{u}})|^{2} \right) \right\rangle_{\Pi_{|\psi\rangle}(\mathbf{u})}
\end{equation}
for $n=1$, where $\langle ... \rangle_{\Pi_{|\psi\rangle}(\mathbf{u})}$ is the average over $\Pi_{|\psi\rangle}(\mathbf{u})$ obtained with sampling. Note however that the number of samples required to estimate $\mathcal{M}_n$ within a given error scales polynomially with $N$ only for $n=1$. For $n \neq 1$, including our case of interest ($n=1/2$), the number of samples required grows exponentially with $N$. This exponential scaling becomes a significant bottleneck for studying large systems.

In contrast, the thermodynamics integration approach offers a more efficient strategy. This method only requires computing the expectation value  $\left\langle \log|\Tilde{W}_{\rho}(\mathbf{u})| \right\rangle_\beta $ for different values of $\beta$ from $\beta=1$ to $\beta=2$ (see Eq. \eqref{eq:mana_thermo}). Importantly, the variance of $\log|\Tilde{W}_{\rho}(\mathbf{u})|$ only scales polynomially with $N$ \cite{deBoer2019}, making its estimation efficient for any value of $\beta$. This technique thus circumvents the exponentially difficult task of estimating the mana, enabling to study mana in large systems.

\section{Quantum Potts model} \label{sec:potts}
In this work, we consider the quantum Potts model, which can be seen as the generalization of the quantum Ising model with $d$ states per site \cite{Wu1982}. The Hamiltonian is
given by 
\begin{equation} \label{eq:ham_potts}
    H_{\text{Potts}}=-J \sum_{\langle i,j \rangle} \sum_{k=1}^{d-1} X_i^k X_j^{d-k}  -h \sum_i \sum_{k=1}^{d-1} Z_i^k ,
\end{equation}
 where $X,Z$ are the shift and clock operators in Eq. \eqref{eq:clock_and_shift}. Here we focus on the case $d=3$.
 The point $h_c=1$ is a critical self-dual point, which is governed by a CFT for $d\leq 4$. For $d=3$, the central charge is $c=4/5$ in the ferromagnetic case ($J=1$) and $c=1$ in the antiferromagnetic case ($J=-1$) \cite{Affleck1998,DiFrancesco, Lahtinen2021}.

 We will also consider an extension of the Potts model introduced in Ref. \cite{alcaraz2016}. The Hamiltonian is given by
\begin{equation} \label{eq:potts_extension}
\begin{split}
    H_{\text{Potts}}(p)= &- \sum_{\langle i,j \rangle} \sum_{k=1}^{d-1} X_i^k X_j^{d-k}  - \sum_i \sum_{k=1}^{d-1} Z_i^k  \\
    &- p\sum_{\langle\langle i,j \rangle\rangle} \sum_{k=1}^{d-1} X_i^k X_j^{d-k} - p\sum_{\langle i,j \rangle} \sum_{k=1}^{d-1} Z_i^k Z_j^{d-k} .
    \end{split}
\end{equation}
The model is self-dual at any $p$, and the case $p=0$ corresponds to the self-dual point $h=1$ in Eq. \eqref{eq:ham_potts}, which is an integrable point. For $p \neq 0$, the model is not integrable, but it is expected that they are described by the same CFT at $p=0$ for sufficiently small $p$ \cite{alcaraz2016}.

\section{Numerical results}
\label{sec:num_results}

We now present our numerical results on the mana in the quantum Potts model on a periodic chain. We obtain the ground state using TTN ground state variational search algorithm \cite{gerster2014,Silvi2019}, and then we sample the discrete Wigner function of the ground state using Monte Carlo sampling on TTN discussed in Ref. \cite{tarabunga2023manybody}. We use  bond dimension up to $\chi=36$.  
Here, we compute the full-state mana using Eq. \eqref{eq:mana_thermo}, while the mutual mana is evaluated using Eq. \eqref{eq:estimator_mutual_mana}.

The mana density is shown in Fig. \ref{fig:mana}a. We observe that $\mathcal{M}/L$ reaches a maximum at the critical point $h_c=1$, which confirm the results of Ref. \cite{white2021}. More importantly, with the large systems we are able to simulate, we obtain good data collapse, shown in Fig. \ref{fig:mana}b. Overall, these results are also similar to the behavior of SEs, which are studied for $n\in \{1,2 \}$ in Ref. \cite{tarabunga2023manybody}. Indeed, in this case the mana is identical to the SE with $n=1/2$ through the proposition in Sec.~\ref{sec:mana_stabilizer}~\footnote{
 In the case of three-state Potts model, the ground state satisfies $A_\mathbf{0}| \psi \rangle = | \psi \rangle$, due to the global $S_3$ symmetry of permuting the three $X$ eigenstates. A similar statement holds for all $d$-state Potts model for odd prime $d$.  
}.

\begin{figure} 
    \centering
    \includegraphics[width=1\linewidth]{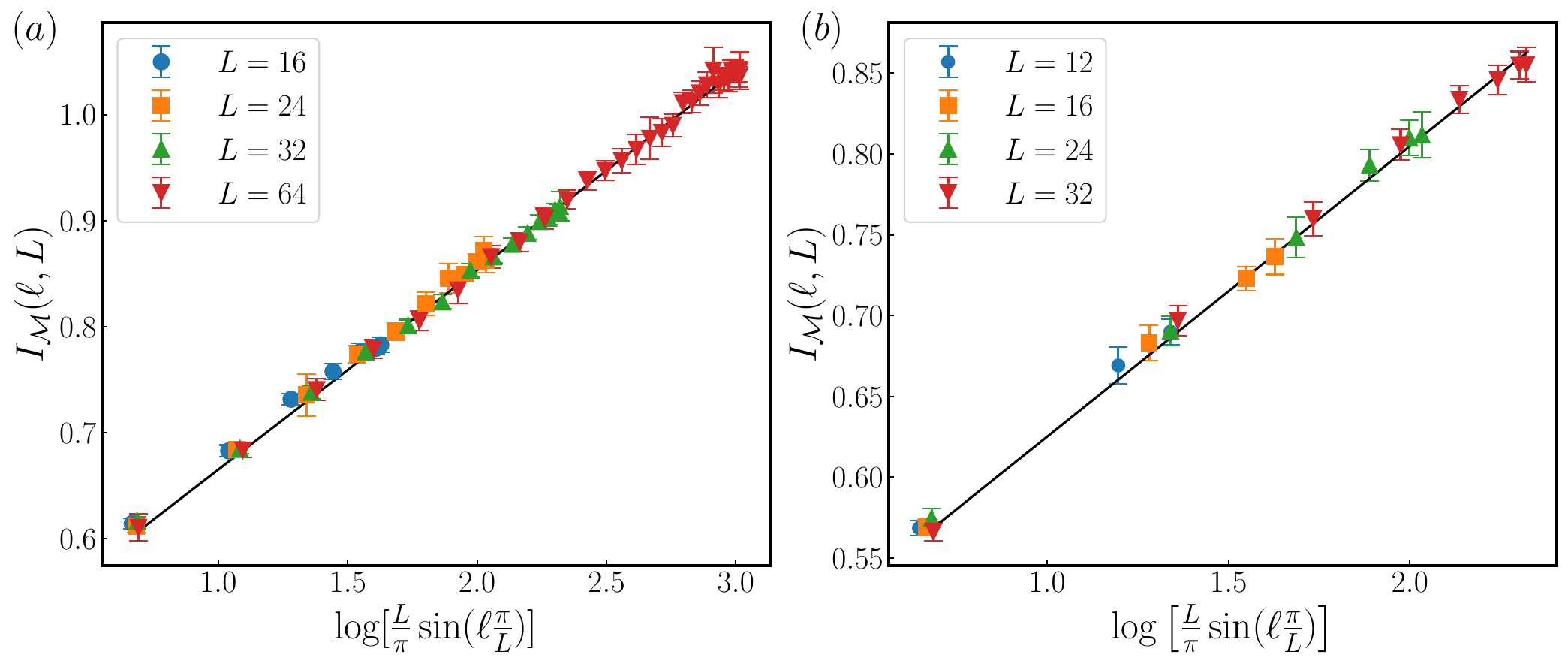}
    \caption{Mutual mana $I_\mathcal{M}(\ell,L)$ in the ground state of the quantum Potts model at the critical point $h/J=1$ with (a) $J=1$ and (b) $J=-1$. The solid line denotes the linear fit obtained for the largest size. Clearly the data of different system sizes collapse in the straight line.  We observe odd-even effects for $J=-1$, and thus we plot only the results for even $\ell$ for clarity.}
    \label{fig:mutual_mana}
\end{figure}
Next, we investigated the scaling of mutual mana (Eq. \eqref{eq:mutual_mana}) at the critical point $h_c=1$. The results are shown in Fig. \ref{fig:mutual_mana}a(b) for $J=1$ ($J=-1$) for sizes up to $L=64$ ($L=32$) . We observe that the mutual mana is approximately proportional to $\log \left[ \frac{L}{\pi} \sin (\ell \frac{\pi}{L}) \right]$, similarly to  the entanglement and Shannon entropy in CFT. 
However, we cannot make a direct connection between the slope and the central charge of the associated CFT \footnote{Actually, there are also disputes  regarding the slope of Shannon mutual information, and whether it is truly equal to $c/4$. See \cite{stephan2014}.}. This is expected since mana is a basis-dependent quantity, and hence the proportionality factor would likely depend on the choice of basis.

\begin{figure} 
    \centering
    \includegraphics[width=1\linewidth]{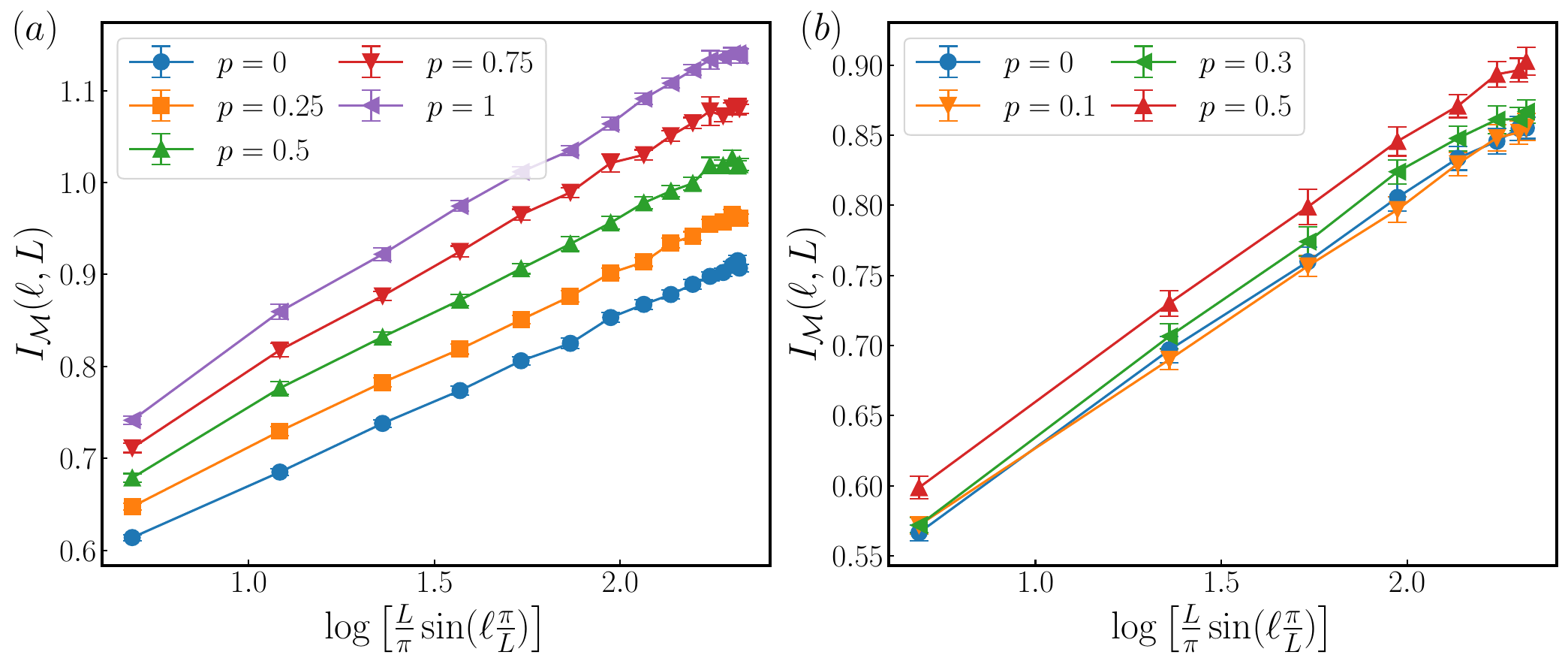}
    \caption{ Mutual mana $I_\mathcal{M}(\ell,L)$  for various values of $p$ in the extension of the quantum Potts model (Eq. \eqref{eq:potts_extension}) with (a) $J=1$ and (b) $J=-1$. The logarithmic scaling is also observed at the non-integrable points $p \neq 0$. The system size is $L=32$.}
    \label{fig:mana_p}
\end{figure}

We now turn to the extension of the Potts model in Eq. \eqref{eq:potts_extension}. Fig. \ref{fig:mana_p} shows the mutual mana for various values of $p$ in a chain of $L=32$ sites. These results clearly reveal a linear scaling of the mutual mana with respect to $\log \left[ \frac{L}{\pi} \sin (\ell \frac{\pi}{L}) \right]$, which holds true even at the non-integrable points. Notably, the slope of the linear growth shows little variation upon increasing $p$. Based on these findings, we conjecture that the slope is universal and determined by the underlying CFT, although possibly not by a simple relation with central charge as entanglement and Shannon entropy.

\begin{figure} 
    \centering
    \includegraphics[width=1\linewidth]{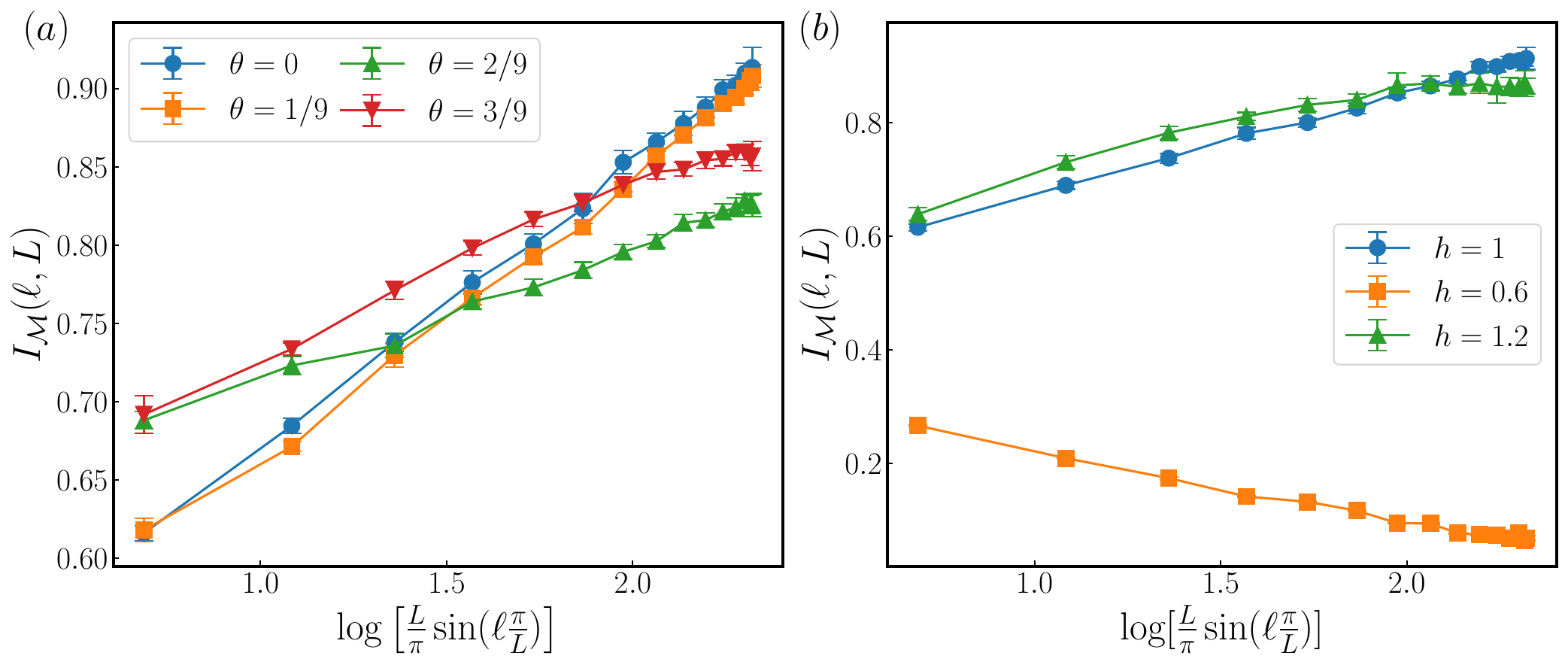}
    \caption{ (a) Mutual mana $I_\mathcal{M}(\ell,L)$ after performing the unitary transformation $T^{\otimes N}_\theta$, where $T_\theta = \text{diag}(1,e^{i\theta},e^{-i\theta})$, to the ground state at $h=1$ and $J=1$. (b) Mutual mana $I_\mathcal{M}(\ell,L)$ in the ground state of the three-state Potts model and at three different transverse field strength. The logarithmic scaling is only observed at the critical point at $h=1$. In contrast, the mutual mana saturates both at $h>1$ and $h<1$. The system size is $L=32$.}
    \label{fig:offcrit}
\end{figure}

Since mana depends on the chosen basis, an important question is whether or not the logarithmic scaling persists under local basis change. To address this question, we show the mutual mana after applying unitary transformation $T^{\otimes N}_\theta$, where $T_\theta = \text{diag}(1,e^{i\theta},e^{-i\theta})$, to the ground state at $h=1$ in Fig. \ref{fig:offcrit}a. Note that $\theta=2/9$ corresponds to the canonical $T$-gate for qutrit. We see that the logarithmic scaling remains evident up to $\theta=2/9$, while it becomes less apparent for $\theta=3/9$, possibly due to finite-size effects.

Finally, in order to contrast with the behavior away from criticality, we plot the scaling of mutual mana both at and away from the critical point in Fig. \ref{fig:offcrit}b. We see that the logarithmic scaling is observed only at the critical point, while away from the critical point the mutual mana saturates at large $\ell$. 

\section{Conclusions and outlook} \label{sec:conclusions}
In this work, we investigate the behavior of mana around criticality in quantum Potts models and its extension. We introduce Rényi version of mana, which enables us to calculate mana for large system sizes. Our results on mutual mana provide clear evidence of logarithmic scaling with distance in CFT, while it reaches saturation in gapped phases. This illustrates that, much like entanglement, the scaling of mutual mana provides a means to distinguish between critical and non-critical behaviors. Moreover, our results on the non-integrable extension indicate the universal character of the logarithmic scaling at criticality. Combined with the findings of recent studies indicating that non-stabilizerness is considerably less susceptible to errors arising from finite bond dimensions \cite{haug2023quantifying,tarabunga2023manybody,frau2024nonstabilizerness}, our work highlights the potential of non-stabilizerness as a useful tool to detect and characterize conformally invariant quantum chains, particularly in the context of tensor network simulations. 

Our work opens up many interesting directions for future investigations. Although mana is only defined for odd prime local dimension, several possible extensions have been proposed for qubits \cite{delfosse2015, Howard2014,vega2017,kocia2017,raussendorf2020}. It would be interesting to employ them to investigate the qubit case, in particular regarding its scaling in CFT. A more comprehensive examination of mutual mana in CFT also warrants futher investigation, for instance by looking at different partitioning schemes. Additionally, it would be interesting to study the behavior of mana minimized over all possible bases.

Furthermore, our methods enable the exploration of mana in various scenarios, such as quench dynamics \cite{sewell2022,goto2022}, open systems and finite-temperature scenarios. In addition, it would be interesting to adapt our approach in different classes of tensor network states such as PEPS \cite{cirac2021}  to investigate the mana in higher dimensions. Another interesting direction is to systematically study and compare the behavior of mana entropy and stabilizer entropy, which may provide insights into how to construct a genuine measure of non-stabilizerness for qubits that is efficiently computable. Finally, while here the mana entropy is introduced to facilitate the numerical computations of mana, it may also be helpful in the analytical investigation of mana in important classes of states, such as the quantum hypergraph states \cite{chen2023magic}.

 \begin{acknowledgments}
We thank M. Dalmonte, E. Tirrito, T. Chanda, C. Castelnovo for insightful discussions  and collaborations on related topics.
This work was partly supported by the PNRR MUR project PE0000023-NQSTI, and by the EU-Flagship programme Pasquans2.
 We acknowledge support from the Simons Foundation through Award 284558FY19 to the ICTP. 

 Our TTN codes have been implemented using C++ Itensor library~\cite{itensor22}.

\end{acknowledgments}

\appendix

\section{Proof of proposition} \label{sec:proof}
We consider a system of $N$ $d$-level sites, where each sites can take values $\sigma_i \in \{0,1,...,{d-1} \}$. Consider a state $| \psi \rangle = \sum_{{\boldsymbol{\sigma}}} c_{\boldsymbol{\sigma}} | \boldsymbol{\sigma} \rangle $, where $c_{\boldsymbol{\sigma}}  = \langle \boldsymbol{\sigma} | \psi \rangle $ and $\boldsymbol{\sigma}$ is the configuration of the system, $\boldsymbol{\sigma}=\{\sigma_1,...,\sigma_N \}$.

In the main text, we state the following proposition:

{\bf Proposition:} { \em Let $| \psi \rangle$ be an $N$-qudit pure state. If $A_\mathbf{b}$ is a phase-space operator such that $A_\mathbf{b} | \psi \rangle = \lambda | \psi \rangle$, where $\lambda \in \{ +1,-1\}$, then
\begin{equation}
    \lambda \langle \psi |  A_{\mathbf{a}+\mathbf{b}}   | \psi \rangle = \langle \psi |   T_{2\mathbf{a}}   | \psi \rangle \omega^{2(\mathbf{b}\mathbf{a'}-\mathbf{b'}\mathbf{a})}
\end{equation}
for all $\mathbf{a} \in \mathbb{Z}_d^{2N}$.}

We will first prove that the following equation holds:
\begin{equation} \label{eq:wigner_pauli}
    A_\mathbf{a+b} A_\mathbf{b} = T_{2\mathbf{a}} \omega^{2(\mathbf{b}\mathbf{a'}-\mathbf{b'}\mathbf{a})}
\end{equation}

Firstly, we note that $A_\mathbf{a}$ can be written as
\begin{equation}
    A_\mathbf{a} = \frac{1}{d^N} \bigotimes_{i=1}^{N} \sum_{b,b'} \omega^{a_i b'- a_i' b} T_{b,b'} 
\end{equation}

The action of $A_\mathbf{a}$ on a basis state $| \sigma \rangle$ is
\begin{equation} \label{eq:action_wigner}
\begin{split}
     A_\mathbf{a} | \boldsymbol{\sigma} \rangle &= \frac{1}{d^N}  \prod_{i=1}^{N} \sum_{b_i,b_i'} \omega^{a_i b_i'- a_i' b_i} T_{b_i,b_i'} | \sigma_i \rangle \\
    &= \frac{1}{d^N}  \prod_{i=1}^{N} \sum_{b_i,b_i'} \omega^{a_i b_i'- a_i' b_i} \omega^{-2^{-1}b_i b_i'} Z^{b_i} X^{b_i'}| \sigma_i \rangle \\
    &= \frac{1}{d^N}  \prod_{i=1}^{N} \sum_{b_i,b_i'} \omega^{a_i b_i'- a_i' b_i+b_i(\sigma_i+2^{-1} b_i') }  | \sigma_i +b_i' \rangle \\
    &=  \prod_{i=1}^{N} \sum_{b_i'} \omega^{a_i b_i'} \delta_{\sigma_i+2^{-1}b_i'-a_i',0}  | \sigma_i +b_i' \rangle \\
    &=   \omega^{2\mathbf{a} (\mathbf{a}'-\boldsymbol{\sigma})}   | 2\mathbf{a}'-\boldsymbol{\sigma} \rangle. \\
\end{split}
\end{equation}

On the other hand, the action of $T_\mathbf{a}$ is
\begin{equation} \label{eq:action_pauli}
\begin{split}
    T_\mathbf{a} | \boldsymbol{\sigma} \rangle &=   \prod_{i=1}^{N}   T_{a_i,a_i'} | \sigma_i \rangle \\
    &=   \prod_{i=1}^{N}   \omega^{-2^{-1}a_ia_i'} Z^{a_i} X^{a_i'}| \sigma_i \rangle \\
    &=    \omega^{\mathbf{a} .(2^{-1}\mathbf{a}'+\boldsymbol{\sigma}) } | \mathbf{a}'+\boldsymbol{\sigma} \rangle .  \\
\end{split}
\end{equation}

From Eq. \eqref{eq:action_wigner}, we have that $A_\mathbf{0} | \boldsymbol{\sigma} \rangle=| -\boldsymbol{\sigma} \rangle $. Then,
\begin{equation} \label{eq:A_a-A_0}
\begin{split} 
    A_\mathbf{a} A_\mathbf{0} | \boldsymbol{\sigma} \rangle &= A_\mathbf{a}  | -\boldsymbol{\sigma} \rangle =    \omega^{2\mathbf{a} (\mathbf{a}'+\boldsymbol{\sigma})}   | 2\mathbf{a}'+\boldsymbol{\sigma} \rangle = T_{2\mathbf{a}} | \boldsymbol{\sigma} \rangle.
\end{split}
\end{equation}
Since Eq. \eqref{eq:A_a-A_0} holds for all basis states $| \boldsymbol{\sigma} \rangle$, then $A_\mathbf{a} A_\mathbf{0} = T_{2\mathbf{a}}$. This proves Eq. \eqref{eq:wigner_pauli} in the case $\mathbf{b}=\mathbf{0}$.

Now, using $A_{\mathbf{a}}=T_{\mathbf{a}} A_{\mathbf{0}}T_{\mathbf{a}}^\dagger$ and the commutation relation $T_{\mathbf{a}}T_{\mathbf{b}}=\omega^{\mathbf{a}\mathbf{b'}-\mathbf{a'}\mathbf{b}} T_{\mathbf{b}}T_{\mathbf{a}}$, we have
\begin{equation}
    \begin{split}
        A_{\mathbf{a}+\mathbf{b}} A_\mathbf{b} &= A_{\mathbf{a}+\mathbf{b}} T_{\mathbf{b}} A_\mathbf{0} T_{\mathbf{b}}^\dagger \\
        &= T_{\mathbf{b}} T_{\mathbf{b}}^\dagger A_{\mathbf{a}+\mathbf{b}} T_{\mathbf{b}} A_\mathbf{0} T_{\mathbf{b}}^\dagger \\
        &= T_{\mathbf{b}} A_{\mathbf{a}}  A_\mathbf{0} T_{\mathbf{b}}^\dagger \\
        &= T_{\mathbf{b}} T_{2\mathbf{a}}   T_{\mathbf{b}}^\dagger \\
        &=  T_{2\mathbf{a}}    \omega^{2(\mathbf{b}\mathbf{a'}-\mathbf{b'}\mathbf{a})}.\\
    \end{split}
\end{equation}
This concludes the proof of Eq. \eqref{eq:wigner_pauli}.

The proposition now immediately follows as a corollary of Eq. \eqref{eq:wigner_pauli}. Indeed, if $A_\mathbf{b} | \psi \rangle = \lambda | \psi \rangle$, then
\begin{equation}
    \lambda \langle \psi |  A_{\mathbf{a}+\mathbf{b}}   | \psi \rangle =  \langle \psi |  A_{\mathbf{a}+\mathbf{b}} A_{\mathbf{b}}   | \psi \rangle = \langle \psi |   T_{2\mathbf{a}}   | \psi \rangle \omega^{2(\mathbf{b}\mathbf{a'}-\mathbf{b'}\mathbf{a})} \qed 
\end{equation}

\section{Relations with other non-stabilizerness monotones} \label{sec:relations}
Here we discuss the relations between the mana entropy, the min-relative entropy, the free robustness of magic, and the stabilizer nullity, restricting to the case of pure states.  

\subsection{Relation with the free robustness of magic}
We denote by STAB the set of all stabilizer states.
The free robustness of magic is defined as
\begin{equation}
    \mathcal{R}(\rho) = \min s \quad \text{s.t.} \quad   \rho = (1+s) \sigma - s\sigma', \sigma,\sigma' \in \text{STAB}
\end{equation}
while the min-relative entropy is defined as
\begin{equation}
    \mathcal{D}_{min}(|\psi \rangle) = -\log F_{\text{STAB}}(|\psi \rangle)
\end{equation}
where $F_{STAB}$ is the stabilizer fidelity defined as
\begin{equation}
    F_{STAB}(|\psi \rangle) = \max_{|\phi \rangle \in \text{STAB}} | \langle \phi |\psi \rangle|^2.
\end{equation}

In \cite{liu2022}, it was shown that
\begin{equation}
    \mathcal{M}(\rho ) \leq \log \left[ 2 \mathcal{R}(\rho )+1 \right] .
\end{equation}
Notice that $\mathcal{M}_{1/2}=2\mathcal{M}$, so that we have by the hierarchy of Renyi entropies
\begin{equation}
    \mathcal{M}_{n}(|\psi \rangle) \leq 2\log \left[2\mathcal{R}(|\psi \rangle)+1  \right] \quad (n\geq1/2) .
\end{equation}

\subsection{Relation with min-relative entropy of magic}
Next, we will show the following inequality holds
\begin{equation} \label{eq:dmin}
     \mathcal{M}_{n}(|\psi \rangle) \leq \frac{2n}{n-1} \mathcal{D}_{min}(|\psi \rangle) \quad (n >1) .
\end{equation}
The proof we give below is inspired by the proof of similar inequality for SE given in Ref. \cite{haug2023stabilizer}. 

Given a pure stabilizer state $| \phi \rangle$, we denote $\mathcal{U}(| \phi \rangle):=\{ \mathbf{u}\in \mathbb{Z}_d^{2N} : W_{| \phi \rangle}(\mathbf{u})=1 \}$. By the discrete Hudson's theorem \cite{Gross2006}, we have $|\mathcal{U}(| \phi \rangle)|=d^N$, and moreover  $W_{| \phi \rangle}(\mathbf{u})=0$ for $\mathbf{u}\notin \mathcal{U}(| \phi \rangle)$. Thus, we can write
\begin{equation}
    | \phi \rangle \langle \phi | = \frac{1}{d^N}  \sum_{\mathbf{u} \in \mathcal{U}(| \phi \rangle)} A_\mathbf{u}.
\end{equation}

We have
\begin{equation*} 
    \begin{split}
        \frac{1}{d^N} \sum_{\mathbf{u}} |\langle \psi | A_\mathbf{u}| \psi \rangle|^{2n} &\geq \frac{1}{d^N} \sum_{\mathbf{u} \in \mathcal{U}(| \phi \rangle)} |\langle \psi | A_\mathbf{u}| \psi \rangle|^{2n} \\
        & \geq \frac{1}{d^{2nN}} \left( \sum_{\mathbf{u} \in \mathcal{U}(| \phi \rangle)} |\langle \psi | A_\mathbf{u}| \psi \rangle| \right)^{2n} \\
        & \geq \frac{1}{d^{2nN}} \left| \sum_{\mathbf{u} \in \mathcal{U}(| \phi \rangle)} \langle \psi | A_\mathbf{u}| \psi \rangle \right|^{2n} \\
        & = \frac{1}{d^{2nN}} \left| \langle \psi | \phi \rangle \right|^{4n} \\
    \end{split}
\end{equation*}
Taking logarithm on both sides and dividing by $(n-1)$, we have
\begin{equation}
    \mathcal{M}_{n}(|\psi \rangle) \leq \frac{2n}{n-1} \log \left| \langle \psi | \phi \rangle \right|^{2}  \quad (n >1)
\end{equation}

Minimizing the right hand side, we finally obtain Eq. \eqref{eq:dmin}. 

\begin{figure} 
    \centering
    \includegraphics[width=0.8\linewidth]{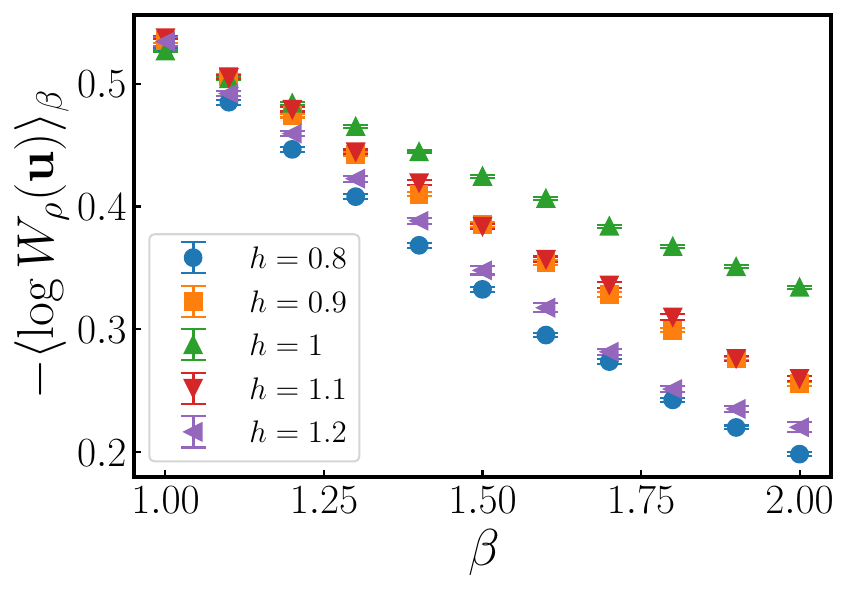}
    \caption{$- \left\langle \log (\Tilde{W}_\rho(\mathbf{u}))\right\rangle_\beta$ as a function of $\beta$ from $\beta=1$ to $\beta=2$. }
    \label{fig:integral}
\end{figure}

\subsection{Relation with stabilizer nullity}
Finally, it can be shown that $\mathcal{M}_{n}(\ket{\psi})$ is related to the stabilizer nullity $\nu(|\psi \rangle)$ \cite{Beverland2020} by the following inequality:
\begin{equation} \label{eq:relation_nullity}
    \mathcal{M}_{n}(\ket{\psi}) \leq \nu(|\psi \rangle) .
\end{equation}

To prove this, we use the known fact that for a state  with stabilizer nullity $\nu(|\psi \rangle)$, there exists a Clifford unitary $C$ such that $C\ket{\psi} = \ket{0}^{N-\nu} \ket{\phi}$, where $\ket{\phi}$ is a pure state of $\nu$ qubits.
Therefore,
\begin{equation}
    \mathcal{M}_n(\ket{\psi}) = \mathcal{M}_n(\ket{0}^{N-\nu} \ket{\phi}) = \mathcal{M}_n(\ket{\phi}) \leq \nu . 
\end{equation}

\section{Numerical integration} \label{sec:integration}
Fig. \ref{fig:integral} shows $- \left\langle \log (\Tilde{W}_\rho(\mathbf{u}))\right\rangle_\beta$ from $\beta=1$ to $\beta=2$. The quantity is integrated using trapezoid rule to give the mana presented in the main text. We see that in all cases considered, the integrand is close to being linear, such that a small number of grids is sufficient to compute the mana with small discretization error. We have checked that increasing the number of grids yields the values of mana that agree within error bars.

\vspace{2ex}\noindent

\bibliographystyle{quantum}
\bibliography{biblio}

\begin{thebibliography}{10}

\bibitem{horodecki2009quantum}
Ryszard Horodecki, Pawe\l{} Horodecki, Micha\l{} Horodecki, and Karol
  Horodecki.
\newblock ``Quantum entanglement''.
\newblock \href{https://dx.doi.org/10.1103/RevModPhys.81.865}{Rev. Mod. Phys.
  {\bf 81}, 865--942}~(2009).

\bibitem{vedral1997quantifying}
V.~Vedral, M.~B. Plenio, M.~A. Rippin, and P.~L. Knight.
\newblock ``Quantifying entanglement''.
\newblock \href{https://dx.doi.org/10.1103/PhysRevLett.78.2275}{Phys. Rev.
  Lett. {\bf 78}, 2275--2279}~(1997).

\bibitem{nielsen2002quantum}
Michael~A. Nielsen and Isaac~L. Chuang.
\newblock ``Quantum computation and quantum information''.
\newblock \href{https://dx.doi.org/10.1017/cbo9780511976667}{Cambridge
  University Press}. ~(2012).

\bibitem{amico2008}
Luigi Amico, Rosario Fazio, Andreas Osterloh, and Vlatko Vedral.
\newblock ``Entanglement in many-body systems''.
\newblock \href{https://dx.doi.org/10.1103/RevModPhys.80.517}{Rev. Mod. Phys.
  {\bf 80}, 517--576}~(2008).

\bibitem{eisert2010}
J.~Eisert, M.~Cramer, and M.~B. Plenio.
\newblock ``Colloquium: Area laws for the entanglement entropy''.
\newblock \href{https://dx.doi.org/10.1103/RevModPhys.82.277}{Rev. Mod. Phys.
  {\bf 82}, 277--306}~(2010).

\bibitem{Holzhey1994}
Christoph Holzhey, Finn Larsen, and Frank Wilczek.
\newblock ``Geometric and renormalized entropy in conformal field theory''.
\newblock \href{https://dx.doi.org/10.1016/0550-3213(94)90402-2}{Nuclear
  Physics B {\bf 424}, 443--467}~(1994).

\bibitem{Calabrese2004}
Pasquale Calabrese and John Cardy.
\newblock ``Entanglement entropy and quantum field theory''.
\newblock \href{https://dx.doi.org/10.1088/1742-5468/2004/06/p06002}{Journal of
  Statistical Mechanics: Theory and Experiment {\bf 2004}, P06002}~(2004).

\bibitem{gottesman1997stabilizer}
Daniel Gottesman.
\newblock ``{Stabilizer Codes and Quantum Error Correction}''~(1997)
  \href{http://arxiv.org/abs/quant-ph/}{arXiv:quant-ph/970505}.

\bibitem{gottesman1998theory}
Daniel Gottesman.
\newblock ``Theory of fault-tolerant quantum computation''.
\newblock \href{https://dx.doi.org/10.1103/PhysRevA.57.127}{Phys. Rev. A {\bf
  57}, 127--137}~(1998).

\bibitem{gottesman1998heisenberg}
Daniel Gottesman.
\newblock ``{The Heisenberg representation of quantum computers}''~(1998)
  \href{http://arxiv.org/abs/quant-ph/}{arXiv:quant-ph/980700}.

\bibitem{aaronson2004improved}
Scott Aaronson and Daniel Gottesman.
\newblock ``Improved simulation of stabilizer circuits''.
\newblock \href{https://dx.doi.org/10.1103/PhysRevA.70.052328}{Phys. Rev. A
  {\bf 70}, 052328}~(2004).

\bibitem{bravyi2005UniversalQuantumComputation}
Sergey Bravyi and Alexei Kitaev.
\newblock ``{Universal quantum computation with ideal Clifford gates and noisy
  ancillas}''.
\newblock \href{https://dx.doi.org/10.1103/PhysRevA.71.022316}{Phys. Rev. A
  {\bf 71}, 022316}~(2005).

\bibitem{bravyi2012magic}
Sergey Bravyi and Jeongwan Haah.
\newblock ``Magic-state distillation with low overhead''.
\newblock \href{https://dx.doi.org/10.1103/PhysRevA.86.052329}{Phys. Rev. A
  {\bf 86}, 052329}~(2012).

\bibitem{campbell2017roads}
Earl~T. Campbell, Barbara~M. Terhal, and Christophe Vuillot.
\newblock ``Roads towards fault-tolerant universal quantum computation''.
\newblock \href{https://dx.doi.org/10.1038/nature23460}{Nature {\bf 549},
  172--179}~(2017).

\bibitem{harrow2017quantum}
Aram~W. Harrow and Ashley Montanaro.
\newblock ``Quantum computational supremacy''.
\newblock \href{https://dx.doi.org/10.1038/nature23458}{Nature {\bf 549},
  203--209}~(2017).

\bibitem{chitambar2019}
Eric Chitambar and Gilad Gour.
\newblock ``Quantum resource theories''.
\newblock \href{https://dx.doi.org/10.1103/RevModPhys.91.025001}{Rev. Mod.
  Phys. {\bf 91}, 025001}~(2019).

\bibitem{white2021}
Christopher~David White, ChunJun Cao, and Brian Swingle.
\newblock ``Conformal field theories are magical''.
\newblock \href{https://dx.doi.org/10.1103/PhysRevB.103.075145}{Phys. Rev. B
  {\bf 103}, 075145}~(2021).

\bibitem{Sarkar2020}
S~Sarkar, C~Mukhopadhyay, and A~Bayat.
\newblock ``Characterization of an operational quantum resource in a critical
  many-body system''.
\newblock \href{https://dx.doi.org/10.1088/1367-2630/aba919}{New Journal of
  Physics {\bf 22}, 083077}~(2020).

\bibitem{oliviero2022ising}
Salvatore F.~E. Oliviero, Lorenzo Leone, and Alioscia Hamma.
\newblock ``Magic-state resource theory for the ground state of the
  transverse-field ising model''.
\newblock \href{https://dx.doi.org/10.1103/PhysRevA.106.042426}{Phys. Rev. A
  {\bf 106}, 042426}~(2022).

\bibitem{haug2023quantifying}
Tobias Haug and Lorenzo Piroli.
\newblock ``Quantifying nonstabilizerness of matrix product states''.
\newblock \href{https://dx.doi.org/10.1103/PhysRevB.107.035148}{Phys. Rev. B
  {\bf 107}, 035148}~(2023).

\bibitem{tarabunga2023manybody}
Poetri~Sonya Tarabunga, Emanuele Tirrito, Titas Chanda, and Marcello Dalmonte.
\newblock ``Many-body magic via pauli-markov chains---from criticality to gauge
  theories''.
\newblock \href{https://dx.doi.org/10.1103/PRXQuantum.4.040317}{PRX Quantum
  {\bf 4}, 040317}~(2023).

\bibitem{frau2024nonstabilizerness}
M.~Frau, P.~S. Tarabunga, M.~Collura, M.~Dalmonte, and E.~Tirrito.
\newblock ``Nonstabilizerness versus entanglement in matrix product states''.
\newblock \href{https://dx.doi.org/10.1103/physrevb.110.045101}{Physical Review
  B{\bf 110}}~(2024).

\bibitem{tirrito2023}
Emanuele Tirrito, Poetri~Sonya Tarabunga, Gugliemo Lami, Titas Chanda, Lorenzo
  Leone, Salvatore F.~E. Oliviero, Marcello Dalmonte, Mario Collura, and
  Alioscia Hamma.
\newblock ``Quantifying nonstabilizerness through entanglement spectrum
  flatness''.
\newblock \href{https://dx.doi.org/10.1103/physreva.109.l040401}{Physical
  Review A{\bf 109}}~(2024).

\bibitem{turkeshi2023measuring}
Xhek Turkeshi, Marco Schir\`o, and Piotr Sierant.
\newblock ``Measuring nonstabilizerness via multifractal flatness''.
\newblock \href{https://dx.doi.org/10.1103/PhysRevA.108.042408}{Phys. Rev. A
  {\bf 108}, 042408}~(2023).

\bibitem{leone2022stabilizer}
Lorenzo Leone, Salvatore F.~E. Oliviero, and Alioscia Hamma.
\newblock ``Stabilizer r\'enyi entropy''.
\newblock \href{https://dx.doi.org/10.1103/PhysRevLett.128.050402}{Phys. Rev.
  Lett. {\bf 128}, 050402}~(2022).

\bibitem{Um2012}
Jaegon Um, Hyunggyu Park, and Haye Hinrichsen.
\newblock ``Entanglement versus mutual information in quantum spin chains''.
\newblock \href{https://dx.doi.org/10.1088/1742-5468/2012/10/p10026}{Journal of
  Statistical Mechanics: Theory and Experiment {\bf 2012}, P10026}~(2012).

\bibitem{alcaraz2013}
F.~C. Alcaraz and M.~A. Rajabpour.
\newblock ``Universal behavior of the shannon mutual information of critical
  quantum chains''.
\newblock \href{https://dx.doi.org/10.1103/PhysRevLett.111.017201}{Phys. Rev.
  Lett. {\bf 111}, 017201}~(2013).

\bibitem{stephan2014}
Jean-Marie St\'ephan.
\newblock ``Shannon and r\'enyi mutual information in quantum critical spin
  chains''.
\newblock \href{https://dx.doi.org/10.1103/PhysRevB.90.045424}{Phys. Rev. B
  {\bf 90}, 045424}~(2014).

\bibitem{alcaraz2015}
F.~C. Alcaraz and M.~A. Rajabpour.
\newblock ``Generalized mutual information of quantum critical chains''.
\newblock \href{https://dx.doi.org/10.1103/PhysRevB.91.155122}{Phys. Rev. B
  {\bf 91}, 155122}~(2015).

\bibitem{alcaraz2015_2}
F.~C. Alcaraz and M.~A. Rajabpour.
\newblock ``Universal behavior of the shannon and r\'enyi mutual information of
  quantum critical chains''.
\newblock \href{https://dx.doi.org/10.1103/PhysRevB.90.075132}{Phys. Rev. B
  {\bf 90}, 075132}~(2014).

\bibitem{alcaraz2016}
F.~C. Alcaraz.
\newblock ``Universal behavior of the shannon mutual information in
  nonintegrable self-dual quantum chains''.
\newblock \href{https://dx.doi.org/10.1103/PhysRevB.94.115116}{Phys. Rev. B
  {\bf 94}, 115116}~(2016).

\bibitem{haug2023stabilizer}
Tobias Haug and Lorenzo Piroli.
\newblock ``Stabilizer entropies and nonstabilizerness monotones''.
\newblock \href{https://dx.doi.org/10.22331/q-2023-08-28-1092}{Quantum {\bf 7},
  1092}~(2023).

\bibitem{tarabunga2023magic}
Poetri~Sonya Tarabunga and Claudio Castelnovo.
\newblock ``Magic in generalized rokhsar-kivelson wavefunctions''.
\newblock \href{https://dx.doi.org/10.22331/q-2024-05-14-1347}{Quantum {\bf 8},
  1347}~(2024).

\bibitem{lópez2024exact}
Jordi Arnau~Montañà López and Pavel Kos.
\newblock ``Exact solution of long-range stabilizer r\'enyi entropy in the
  dual-unitary xxz model''~(2024).
\newblock  \href{http://arxiv.org/abs/2405.04448}{arXiv:2405.04448}.

\bibitem{Veitch2012}
Victor Veitch, Christopher Ferrie, David Gross, and Joseph Emerson.
\newblock ``Negative quasi-probability as a resource for quantum computation''.
\newblock \href{https://dx.doi.org/10.1088/1367-2630/14/11/113011}{New Journal
  of Physics {\bf 14}, 113011}~(2012).

\bibitem{Veitch2014}
Victor Veitch, S~A~Hamed Mousavian, Daniel Gottesman, and Joseph Emerson.
\newblock ``The resource theory of stabilizer quantum computation''.
\newblock \href{https://dx.doi.org/10.1088/1367-2630/16/1/013009}{New Journal
  of Physics {\bf 16}, 013009}~(2014).

\bibitem{sewell2022}
Troy~J. Sewell and Christopher~David White.
\newblock ``Mana and thermalization: Probing the feasibility of near-clifford
  hamiltonian simulation''.
\newblock \href{https://dx.doi.org/10.1103/PhysRevB.106.125130}{Phys. Rev. B
  {\bf 106}, 125130}~(2022).

\bibitem{Hostens2005}
Erik Hostens, Jeroen Dehaene, and Bart De~Moor.
\newblock ``Stabilizer states and clifford operations for systems of arbitrary
  dimensions and modular arithmetic''.
\newblock \href{https://dx.doi.org/10.1103/physreva.71.042315}{Physical Review
  A{\bf 71}}~(2005).

\bibitem{Mari2012}
A.~Mari and J.~Eisert.
\newblock ``Positive wigner functions render classical simulation of quantum
  computation efficient''.
\newblock \href{https://dx.doi.org/10.1103/physrevlett.109.230503}{Physical
  Review Letters{\bf 109}}~(2012).

\bibitem{Wang2020}
Xin Wang, Mark~M. Wilde, and Yuan Su.
\newblock ``Efficiently computable bounds for magic state distillation''.
\newblock \href{https://dx.doi.org/10.1103/physrevlett.124.090505}{Physical
  Review Letters{\bf 124}}~(2020).

\bibitem{Wang2019}
Xin Wang, Mark~M Wilde, and Yuan Su.
\newblock ``Quantifying the magic of quantum channels''.
\newblock \href{https://dx.doi.org/10.1088/1367-2630/ab451d}{New Journal of
  Physics {\bf 21}, 103002}~(2019).

\bibitem{Gross2006}
D.~Gross.
\newblock ``Hudson's theorem for finite-dimensional quantum systems''.
\newblock \href{https://dx.doi.org/10.1063/1.2393152}{Journal of Mathematical
  Physics {\bf 47}, 122107}~(2006).

\bibitem{Wootters1987}
William~K Wootters.
\newblock ``A wigner-function formulation of finite-state quantum mechanics''.
\newblock \href{https://dx.doi.org/10.1016/0003-4916(87)90176-x}{Annals of
  Physics {\bf 176}, 1--21}~(1987).

\bibitem{wigner1932}
E.~Wigner.
\newblock ``On the quantum correction for thermodynamic equilibrium''.
\newblock \href{https://dx.doi.org/10.1103/PhysRev.40.749}{Phys. Rev. {\bf 40},
  749--759}~(1932).

\bibitem{pashayan2015}
Hakop Pashayan, Joel~J. Wallman, and Stephen~D. Bartlett.
\newblock ``Estimating outcome probabilities of quantum circuits using
  quasiprobabilities''.
\newblock \href{https://dx.doi.org/10.1103/PhysRevLett.115.070501}{Phys. Rev.
  Lett. {\bf 115}, 070501}~(2015).

\bibitem{leone2024stabilizer}
Lorenzo Leone and Lennart Bittel.
\newblock ``Stabilizer entropies are monotones for magic-state resource
  theory''~(2024).
\newblock  \href{http://arxiv.org/abs/2404.11652}{arXiv:2404.11652}.

\bibitem{lami2023quantum}
Guglielmo Lami and Mario Collura.
\newblock ``Nonstabilizerness via perfect pauli sampling of matrix product
  states''.
\newblock \href{https://dx.doi.org/10.1103/PhysRevLett.131.180401}{Phys. Rev.
  Lett. {\bf 131}, 180401}~(2023).

\bibitem{tarabunga2024nonstabilizerness}
Poetri~Sonya Tarabunga, Emanuele Tirrito, Mari~Carmen Bañuls, and Marcello
  Dalmonte.
\newblock ``Nonstabilizerness via matrix product states in the pauli basis''.
\newblock \href{https://dx.doi.org/10.1103/physrevlett.133.010601}{Physical
  Review Letters{\bf 133}}~(2024).

\bibitem{deBoer2019}
Jan de~Boer, Jarkko J\"arvel\"a, and Esko Keski-Vakkuri.
\newblock ``Aspects of capacity of entanglement''.
\newblock \href{https://dx.doi.org/10.1103/PhysRevD.99.066012}{Phys. Rev. D
  {\bf 99}, 066012}~(2019).

\bibitem{yao2010}
Hong Yao and Xiao-Liang Qi.
\newblock ``Entanglement entropy and entanglement spectrum of the kitaev
  model''.
\newblock \href{https://dx.doi.org/10.1103/PhysRevLett.105.080501}{Phys. Rev.
  Lett. {\bf 105}, 080501}~(2010).

\bibitem{schliemann2011}
John Schliemann.
\newblock ``Entanglement spectrum and entanglement thermodynamics of quantum
  hall bilayers at $\ensuremath{\nu}=1$''.
\newblock \href{https://dx.doi.org/10.1103/PhysRevB.83.115322}{Phys. Rev. B
  {\bf 83}, 115322}~(2011).

\bibitem{zaletel2011}
Michael~P. Zaletel, Jens~H. Bardarson, and Joel~E. Moore.
\newblock ``Logarithmic terms in entanglement entropies of 2d quantum critical
  points and shannon entropies of spin chains''.
\newblock \href{https://dx.doi.org/10.1103/PhysRevLett.107.020402}{Phys. Rev.
  Lett. {\bf 107}, 020402}~(2011).

\bibitem{Isakov2011}
Sergei~V. Isakov, Matthew~B. Hastings, and Roger~G. Melko.
\newblock ``Topological entanglement entropy of a bose{\textendash}hubbard spin
  liquid''.
\newblock \href{https://dx.doi.org/10.1038/nphys2036}{Nature Physics {\bf 7},
  772--775}~(2011).

\bibitem{block2020}
Matthew~S. Block, Jonathan D'Emidio, and Ribhu~K. Kaul.
\newblock ``Kagome model for a ${\mathbb{z}}_{2}$ quantum spin liquid''.
\newblock \href{https://dx.doi.org/10.1103/PhysRevB.101.020402}{Phys. Rev. B
  {\bf 101}, 020402}~(2020).

\bibitem{Zhao2022}
Jiarui Zhao, Bin-Bin Chen, Yan-Cheng Wang, Zheng Yan, Meng Cheng, and Zi~Yang
  Meng.
\newblock ``Measuring r{\'{e}}nyi entanglement entropy with high efficiency and
  precision in quantum monte carlo simulations''.
\newblock \href{https://dx.doi.org/10.1038/s41535-022-00476-0}{npj Quantum
  Materials{\bf 7}}~(2022).

\bibitem{Silvi2019}
Pietro Silvi, Ferdinand Tschirsich, Matthias Gerster, Johannes J\"{u}nemann,
  Daniel Jaschke, Matteo Rizzi, and Simone Montangero.
\newblock ``The tensor networks anthology: Simulation techniques for many-body
  quantum lattice systems''.
\newblock \href{https://dx.doi.org/10.21468/scipostphyslectnotes.8}{{SciPost}
  Physics Lecture Notes}~(2019).

\bibitem{Wu1982}
F.~Y. Wu.
\newblock ``The potts model''.
\newblock \href{https://dx.doi.org/10.1103/RevModPhys.54.235}{Rev. Mod. Phys.
  {\bf 54}, 235--268}~(1982).

\bibitem{Affleck1998}
Ian Affleck, Masaki Oshikawa, and Hubert Saleur.
\newblock ``Boundary critical phenomena in the three-state potts model''.
\newblock \href{https://dx.doi.org/10.1088/0305-4470/31/28/003}{Journal of
  Physics A: Mathematical and General {\bf 31}, 5827--5842}~(1998).

\bibitem{DiFrancesco}
Philippe Di~Francesco, Pierre Mathieu, and David Sénéchal.
\newblock ``{Conformal field theory}''.
\newblock \href{https://dx.doi.org/10.1007/978-1-4612-2256-9}{Graduate texts in
  contemporary physics}. Springer. New York, NY~(1997).

\bibitem{Lahtinen2021}
Ville Lahtinen, Teresia Mansson, and Eddy Ardonne.
\newblock ``Quantum criticality in many-body parafermion chains''.
\newblock \href{https://dx.doi.org/10.21468/scipostphyscore.4.2.014}{{SciPost}
  Physics Core{\bf 4}}~(2021).

\bibitem{gerster2014}
M.~Gerster, P.~Silvi, M.~Rizzi, R.~Fazio, T.~Calarco, and S.~Montangero.
\newblock ``Unconstrained tree tensor network: An adaptive gauge picture for
  enhanced performance''.
\newblock \href{https://dx.doi.org/10.1103/PhysRevB.90.125154}{Phys. Rev. B
  {\bf 90}, 125154}~(2014).

\bibitem{delfosse2015}
Nicolas Delfosse, Philippe Allard~Guerin, Jacob Bian, and Robert Raussendorf.
\newblock ``Wigner function negativity and contextuality in quantum computation
  on rebits''.
\newblock \href{https://dx.doi.org/10.1103/PhysRevX.5.021003}{Phys. Rev. X {\bf
  5}, 021003}~(2015).

\bibitem{Howard2014}
Mark Howard, Joel Wallman, Victor Veitch, and Joseph Emerson.
\newblock ``Contextuality supplies the `magic' for quantum computation''.
\newblock \href{https://dx.doi.org/10.1038/nature13460}{Nature {\bf 510},
  351--355}~(2014).

\bibitem{vega2017}
Juan Bermejo-Vega, Nicolas Delfosse, Dan~E. Browne, Cihan Okay, and Robert
  Raussendorf.
\newblock ``Contextuality as a resource for models of quantum computation with
  qubits''.
\newblock \href{https://dx.doi.org/10.1103/PhysRevLett.119.120505}{Phys. Rev.
  Lett. {\bf 119}, 120505}~(2017).

\bibitem{kocia2017}
Lucas Kocia and Peter Love.
\newblock ``Discrete wigner formalism for qubits and noncontextuality of
  clifford gates on qubit stabilizer states''.
\newblock \href{https://dx.doi.org/10.1103/PhysRevA.96.062134}{Phys. Rev. A
  {\bf 96}, 062134}~(2017).

\bibitem{raussendorf2020}
Robert Raussendorf, Juani Bermejo-Vega, Emily Tyhurst, Cihan Okay, and Michael
  Zurel.
\newblock ``Phase-space-simulation method for quantum computation with magic
  states on qubits''.
\newblock \href{https://dx.doi.org/10.1103/PhysRevA.101.012350}{Phys. Rev. A
  {\bf 101}, 012350}~(2020).

\bibitem{goto2022}
Kanato Goto, Tomoki Nosaka, and Masahiro Nozaki.
\newblock ``Probing chaos by magic monotones''.
\newblock \href{https://dx.doi.org/10.1103/PhysRevD.106.126009}{Phys. Rev. D
  {\bf 106}, 126009}~(2022).

\bibitem{cirac2021}
J.~Ignacio Cirac, David P\'erez-Garc\'{\i}a, Norbert Schuch, and Frank
  Verstraete.
\newblock ``Matrix product states and projected entangled pair states:
  Concepts, symmetries, theorems''.
\newblock \href{https://dx.doi.org/10.1103/RevModPhys.93.045003}{Rev. Mod.
  Phys. {\bf 93}, 045003}~(2021).

\bibitem{chen2023magic}
Junjie Chen, Yuxuan Yan, and You Zhou.
\newblock ``Magic of quantum hypergraph states''.
\newblock \href{https://dx.doi.org/10.22331/q-2024-05-21-1351}{Quantum {\bf 8},
  1351}~(2024).

\bibitem{itensor22}
Matthew Fishman, Steven~R. White, and E.~Miles Stoudenmire.
\newblock ``{The ITensor Software Library for Tensor Network Calculations}''.
\newblock \href{https://dx.doi.org/10.21468/SciPostPhysCodeb.4}{SciPost Phys.
  CodebasesPage~4}~(2022).

\bibitem{liu2022}
Zi-Wen Liu and Andreas Winter.
\newblock ``Many-body quantum magic''.
\newblock \href{https://dx.doi.org/10.1103/PRXQuantum.3.020333}{PRX Quantum
  {\bf 3}, 020333}~(2022).

\bibitem{Beverland2020}
Michael Beverland, Earl Campbell, Mark Howard, and Vadym Kliuchnikov.
\newblock ``Lower bounds on the non-clifford resources for quantum
  computations''.
\newblock \href{https://dx.doi.org/10.1088/2058-9565/ab8963}{Quantum Science
  and Technology {\bf 5}, 035009}~(2020).

\end{thebibliography}

\end{document}